# New Frontiers in Multidimensional Self-Trapping of Nonlinear Fields and Matter

Yaroslav V. Kartashov[1,2], Gregory E. Astrakharchik[3], Boris A. Malomed[4,5,6,7], and Lluis Torner[1,8]

[1]*ICFO-Institut de Ciencies Fotoniques, The Barcelona Institute of Science and Technology, 08860 Castelldefels (Barcelona), Spain*
[2]*Institute of Spectroscopy, Russian Academy of Sciences, Troitsk, Moscow, 108840, Russia*
[3]*Departamento de Física, Universitat Politècnica de Catalunya, 08034 Barcelona, Spain*
[4]*Department of Physical Electronics, School of Electrical Engineering, Faculty of Engineering, Tel Aviv University, P.O.B. 39040, Ramat Aviv, Tel Aviv, Israel*
[5]*Center for Light-Matter Interaction, Tel Aviv University, POB 39040, Ramat Aviv, Tel Aviv, Israel*
[6]*ITMO University, St. Petersburg 197101, Russia*
[7]*School of Physics and Optoelectronic Engineering, Foshan University, Foshan 528000, China*
[8]*Universitat Politècnica de Catalunya, 08034 Barcelona, Spain*

## Abstract

We review the state of the art and recently obtained theoretical and experimental results for two- and three-dimensional (2D and 3D) solitons and related states, such as quantum droplets, in optical systems, atomic Bose-Einstein condensates (BECs), and other fields – in particular, liquid crystals. The central challenge is avoiding the trend of 2D and 3D solitary states supported by the ubiquitous cubic nonlinearity to be strongly unstable – a property far less present in one-dimensional systems. Many possibilities for the stabilization of multi-dimensional states have been theoretically proposed over the years. Most strategies involve non-cubic nonlinearities or using different sorts of potentials, including periodic ones. New important avenues have arisen recently in systems based on two-component BEC with spin-orbit coupling, which have been predicted to support stable 2D and metastable 3D solitons. An important recent breakthrough is the creation of 3D quantum droplets. These are self-sustained states existing in two-component BECs, which are stabilized by the presence of Lee-Hung-Yang quantum fluctuations around the underlying mean-field states. By and large, multi-dimensional geometries afford unique opportunities to explore the existence of complex self-sustained states, including topologically rich ones, which by their very nature are not possible in one-dimensional geometries. In addition to vortex solitons, these are hopfions, skyrmions, and hybrid vortex-antivortex complexes, which have been predicted in different models. Here we review recent landmark findings in this field and outline outstanding open challenges.

**List of acronyms:**

1D one-dimensional
2D two dimensional
3D three dimensional
BEC Bose-Einstein condensate
FF fundamental frequency
GPE Gross-Pitaevskii equation
GVD group-velocity dispersion
LHY Lee-Huang-Yang (correction to the mean-field dynamics of BEC)
MF mean field
MM mixed mode
NLSE nonlinear Schrödinger equation
SH second harmonic
SOC spin-orbit coupling
SV semi-vortex
TS Townes' soliton
VK Vakhitov-Kolokolov (stability criterion)

**Table of contents:**

# 1. Introduction

Soliton-like bound states, i.e., self-sustained localized packets of waves or matter, are a universal phenomenon known in many areas of physics, including nonlinear photonics (optics, plasmonics, and exciton-polariton condensates), Bose-Einstein condensates and other types of quantum gases, hydrodynamics of classical fluids and superfluids, plasmas, nonlinear dynamical lattices, superconductors, semiconductors, magnetic materials, electronics, hadron matter (considered in terms of the field theory), etcetera [1-3]. In the vast majority of cases, experimental observations of solitons have been reported only in effectively one-dimensional (1D) settings. Higher-dimensional (2D and 3D) geometries offer a variety of novel phenomena – first of all, topological states in the form of vortex rings and vortex tori, i.e., respectively, 2D and 3D solitons with embedded integer vorticity $S$. The wavefront of such states is characterized by the presence of one or several nested phase singularities (in particular, vortex lines in the 3D case), whose topological charge, or winding number, is determined by the total phase $\varphi = 2\pi S$ accumulated in the course of travel along a closed contour surrounding the singularity [4,5]. More sophisticated 3D topological solitons may exist in the form of *hopfions* and *Skyrmions*, that carry two independent topological charges [6]. Such states frequently exhibit closed vortex rings inside the wave field. In particular, hopfions are shaped as twisted toroidal vortex modes, with the phase of the wave function varying both along the vortex ring (which determines the overall vorticity $S$ of the hopfion) and also in a transverse plane, around the ring (which determines the intrinsic twist $m$ of the hopfion). In addition to the more complex structure, multidimensional solitons also often exhibit richer interactions than their 1D counterparts [7].

However, the creation of soliton states in 2D and, especially, in 3D settings is a long-standing challenge to the theory and experiment alike, because even the simplest fundamental solutions of the corresponding nonlinear wave equations are often prone to strong destructive instabilities, such as the onset of the *spatiotemporal collapse* and break-up of fundamental solitons [8-10], and spontaneous azimuthal self-splitting of vortex rings and tori [11,12]. Mechanisms allowing stabilization of higher-dimensional self-trapped states are known in theory, as described below, but they are rare and challenging to realize in practice, to say the least. In particular, in pure optical settings the identification of materials featuring the proper interplay between the dispersion, diffraction, and nonlinearity, required for the creation of *stable, long-living* 3D solitons, alias "light bullets" [13], is still a challenge for the experiment, after these objects were theoretically introduced several decades ago, and studied in detail, also theoretically, since then.

Earlier approaches to the realization of stabilized multidimensional solitons were reviewed in [4]. Some of more recent results were briefly summarized in the update [14] and in [15]. Schemes for the stabilization of 2D and 3D solitons, including solitary vortices, which were theoretically elaborated in early works, were based on the use of non-Kerr nonlinearities, discreteness, and dissipative media, to name the most essential ones. Significant progress has been made in the area during the last years, including several important advances and breakthroughs that have been achieved very recently. In particular, the first examples of stable multidimensional states in ultracold bosonic gases, in the form of quantum droplets, the stability of which is grounded on the Lee-Hung-Yang corrections to the mean-field theory [16], have been reported very recently [17-26]. Further advances may be foreseen in the nearest future, as several new concepts and techniques are currently subjects of intensive investigation.

This review is addressed to researchers working in various areas of physics and aims to stimulate work towards achieving fundamentally novel results. We summarize recent progress in theoretical and experimental investigation of multidimensional trappings, with particular emphasis on 3D states in Optics, atomic Bose-Einstein condensates, and material patterns in liquid crystals. The paper is organized as follows. In section 2 we outline the basic features of multidimensional light trapping and the mechanisms of its occurrence in various uniform optical media, as well as in structured systems, such as photonic lattices and multimode waveguides. In the same section we address manifestations of the formation of spatio-temporal filaments with high-power laser radiation in the ionization regime, and briefly mention 2D solitons in the area of polariton condensates. Section 3 is devoted to the recent theoretical findings for solitons in media with inhomogeneous nonlinearities, which is relevant to both optical and matter-wave realizations. Section 4 describes recent theoretical advances in the elucidation of multidimensional sates in atomic BECs, including matter-wave solitons in optical lattices, stabilization of multidimensional condensates by dint of spin-orbit-coupling, and recent landmark experimental advances in the creation of stable quantum droplets in binary BECs. The topic of 3D topological solitons based on various knotted patterns of the nematic field, generated by laser beams in liquid crystals, is briefly surveyed in section 5. In section 6 we discuss open issues and prospects of further progress.

## 2. Multidimensional states in photonics

### 2.1. Uniform media

Formation of spatiotemporal solitons in dispersive optical media may occur when two fundamental linear effects – group-velocity dispersion (GVD), leading to broadening of the electromagnetic wave packet in time, and diffraction, acting in the transverse directions – are balanced by a focusing nonlinearity, the most universal source of which is provided by the Kerr effect [alias $\chi^{(3)}$ nonlinearity] in dielectric media [1-3]. By and large, a necessary condition for such a balance is anomalous GVD, which implies increasing group velocity with decreasing wavelength, $\lambda$. In the paraxial approximation, the evolution of the wave packet in uniform Kerr-type nonlinear media, is governed by the nonlinear Schrödinger equation (NLSE), also known as Gross-Pitaevskii equation in the context of BEC, for the complex field amplitude $\psi$, written in the normalized form as:

$$i\frac{\partial\psi}{\partial z}=-\frac{1}{2}\left(\frac{\partial^2\psi}{\partial\tau^2}+\frac{\partial^2\psi}{\partial x^2}+\frac{\partial^2\psi}{\partial y^2}\right)-\mathcal{N}(\psi)\psi, \tag{1}$$

where $z$ is the propagation distance, $\tau=t-z/V_{\mathrm{gr}}$ is the reduced time, relative to normal time $t$, $V_{\mathrm{gr}}$ is the group velocity of the carrier wave, $x,y$ are the transverse coordinates, and $\mathcal{N}(\psi)=|\psi|^2$ in Kerr nonlinear media. The same model without the GVD term, $\partial^2\psi/\partial\tau^2$, governs the evolution of 2D beams in the spatial domain.

The NLSE allows the existence of families of bell-shaped (fundamental) spatiotemporal solitons in the form of $\psi = u(\rho)\exp(ikz)$, where $k > 0$ is the propagation constant, $\rho = (x^2 + y^2 + \tau^2)^{1/2}$ is the spatiotemporal radial coordinate, and $u$ is a real function vanishing at $\rho \to \infty$ as $\rho^{-1}\exp[-(2k)^{1/2}\rho]$. The 2D version of such states, with $\rho$ replaced by $r = (x^2 + y^2)^{1/2}$, is known as Townes' soliton (TS) [27], with asymptotic decay $\sim r^{-1/2}\exp[-(2k)^{1/2}r]$ at $r \to \infty$. Historically, TSs were the first species of solitary waves predicted in nonlinear optics, but they cannot be realized experimentally as they are strongly unstable.

The experimental realization of multidimensional solitons faces two main challenges, which have proven to be hard to overcome in spite of focused efforts over the last three decades. First, the states ought to be dynamically stable upon evolution for a given nonlinearity law $\mathcal{N}(\psi)$. Second, in the case of spatiotemporal states, a suitable material setting that simultaneously exhibits high-enough GVD, ultrafast nonlinear response time, and low losses is required, so that diffraction and dispersion effects are comparable without employing too short pulses (ideally, not shorter than a few picoseconds). The latter condition is essential because various higher-order nonlinear effects, which are not included in Eq. (1), dramatically increase as pulses become shorter, preventing the balance required for the stationary propagation of the light bullets.

In focusing media, the stability of fundamental solitons is often determined by the Vakhitov-Kolokolov (VK) stability criterion [28], which predicts instability of the solutions that feature $dE/dk < 0$, and stability under the necessary, but sometimes not sufficient, condition $dE/dk > 0$, where $E = \iiint |\psi|^2 dxdyd\tau$ is the total energy of the spatiotemporal state. It can be readily shown [4] that in uniform Kerr media the scaling relation $E \sim k^{1-D/2}$ between the energy and the propagation constant holds, where $D = 2,3$ is the space dimension. This scaling immediately implies VK instability of 3D solitons. In the critical 2D case, which corresponds to the above-mentioned TSs, $E$ does not depend on $k$. In this case, the instability still persists, becoming nonlinear and developing sub-exponentially, in contrast to the exponential instability growth in the 3D setting. Multidimensional solitons can carry vorticity [29], which makes them additionally vulnerable to instability against azimuthal self-splitting in cubic media.

One possibility to create stable multidimensional solitons, including light bullets, was provided by models with quadratic ($\chi^{(2)}$) nonlinearities, which do support fully stable solutions in 2D and 3D geometries [30-34]. Stable spatial 2D fundamental solitons were experimentally observed in [35], while vortex-carrying solitons are unstable [12], as observed in experiments that showed splitting of the vortex-ring solitons into several fragments [36]. The existence of stable 3D fundamental spatiotemporal solitons in $\chi^{(2)}$ media was originally pointed out in [37], and elaborated in several works afterwards (see, in particular, [38]). 2D spatiotemporal solitons, namely, wave packets with the nonlinearity balancing effective temporal spreading and diffraction in only one transverse spatial direction were created in a series of experiments that employed a specific trick to generate suitable effective anomalous GVD [39,40]. Several potential approaches to generate 3D states, such as using tandem or metamaterial structures with engineered group velocity, GVD, and nonlinearity have been put forward [41]. A possibility of their experimental exploration and demonstration remains open.

Another strategy for the realization of stable multidimensional states relies on the use of materials with saturable or competing nonlinearities, such as $\mathcal{N}(\psi) = |\psi|^2/(1+\nu|\psi|^2)$ or $\mathcal{N}(\psi) = |\psi|^2 - \nu|\psi|^4$ (with real coefficient $\nu > 0$ representing the quintic, alias $\chi^{(5)}$, self-defocusing nonlinearity), leading to new forms of the $E(k)$ dependencies. Saturable nonlinearity is typical for photorefractive crystals, where a nonlinear change of the refractive index is induced via the electro-optic effect (for a review about its application to spatial solitons, see [42]), in vapors of alkali metals [43] – in particular, sodium vapors [44] – and in other media. The use of saturable nonlinearities has made it possible to create a great variety of stable 2D spatial solitons. Among them are solitons supported by the quasi-steady-state [45] and steady-state screening [46,47], that were observed in photorefractive crystals at low power levels. In [48] it was shown theoretically that saturable nonlinearity leads to nonmonotonic $E(k)$ curves, parts of which correspond to stable 3D light bullets.

While standard saturable nonlinearities do not suppress azimuthal instabilities of solitons carrying vorticity [49], competing nonlinearities have been predicted to support stable vortex solitons in both 2D [50,51] and 3D [34,52] geometries. Stability is achieved close to the regime where the contributions from the $\chi^{(3)}$ and $\chi^{(5)}$ nonlinearities almost balance each other, leading to robust flat-top solitons, which, in some sense, may be described as exhibiting "liquid-like" behavior [53]. Formation of 2D fundamental and vortex solitons attributed in part to effective competing $\chi^{(3)}-\chi^{(5)}$ nonlinearities was experimentally reported in $CS_2$ [54,55] and in suspensions of metallic nanoparticles [56], albeit in the presence of significant linear and nonlinear absorption, which severely limits the applicability of such materials in their present form.

Competition of nonlinearities of different orders is one of the mechanisms supporting stable dissipative solitons in models of laser cavities, based on the complex Ginzburg-Landau equation [57,58]. This is an extension of Eq. (1) with $\mathcal{N}(\psi)=i\gamma+(1-i\varepsilon)|\psi|^2-(\nu-i\mu)|\psi|^4$, accounting for linear losses $\gamma$, nonlinear gain $\varepsilon$, and nonlinear absorption $\mu$, and with spectral filtering $\beta$ included into the $(1+i\beta)\partial^2\psi/\partial\tau^2$ expression that replaces the GVD term in Eq. (1). Complex Ginzburg-Landau models give rise to stable 3D fundamental solitons [59,60] and vortex tori [61]. Asymmetric rotating and precessing 3D vortex solitons [62] and vortex knots [63] have been predicted in media with saturable gain and absorption. Recently, dissipative mode-locked cavity solitons with the pulse duration shorter than the cavity round-trip, which may be considered as isolated 3D objects, were observed in a vertical-cavity surface-emitting laser [64]. Such dissipative settings are promising for the generation of multidimensional solitons because they appear as stable attractors with a broad attraction basin, a property that facilitates their excitation.

The possibility of stable multidimensional states has been shown also in nonlocal media [65]. The respective nonlinear nonlocal response in Eq. (1), $\mathcal{N}(\psi)=\int \mathrm{K}(\mathbf{r}'-\mathbf{r})|\psi(\mathbf{r}')|^2\, d\mathbf{r}'$, at a given point $\mathbf{r}$ depends on the entire spatial intensity profile, where $\mathrm{K}(\mathbf{r})$ is the nonlocality kernel that depends on the type of the nonlinearity. The nonlocality can smooth out the spatial intensity fluctuations, and thus prevent the development of the instability driven by the Kerr nonlinearity. Nonlocality is typical, e.g., for reorientational [66] and thermal [67] optical nonlinearities that respond to time-averaged intensities. Such a nonlocal response is typically very slow, thus it is insensitive to the pulsed character of the excitation and can be used for spatial self-trapping of pulse-train solitons, while fast Kerr nonlinearity, which acts only around the maximum of each pulse, may be used for the dispersion compensation [68]. A similar mechanism allowed the recent observation of dark (in time)-bright (in space) 3D pulse train solitons in photorefractive crystals [69].

## 2.2. Photonic lattices

A universal method for the stabilization of both fundamental and vortex solitons in 2D [70-72] and 3D [73-75] geometries with the Kerr focusing nonlinearity is offered by the use of spatially periodic potentials, induced by photonic lattices or arrays [76-79]. Such potentials are created by shallow $(\delta n \sim 10^{-3})$ modulations of the refractive index of the material in the transverse plane. Light propagation in such media is governed by the modified version of Eq. (1):

$$i\frac{\partial\psi}{\partial z}=-\frac{1}{2}\left(\frac{\partial^2\psi}{\partial\tau^2}+\frac{\partial^2\psi}{\partial x^2}+\frac{\partial^2\psi}{\partial y^2}\right)-\mathcal{N}(\psi)\psi-\mathcal{R}(x,y)\psi, \tag{2}$$

where the effective potential, $-\mathcal{R}\sim-\delta n(x,y)$, represents the shape of the periodically modulated refractive-index landscape. Different approaches to the creation of such potentials have been suggested. Two salient techniques are (for other methods see [78,79]). First, the induction of optical lattices in photorefractive media [80], that allowed the observation of 2D fundamental lattice solitons [77], vortex solitons in square [81,82] and hexagonal [83] lattices, soliton trains [84] and dislocations [85]. Second, di-rect laser writing of arrays by femtosecond pulses [86]. Optical induction allows erasing and rewriting lattices in the same sample, while laser writing allows creating permanent periodic waveguiding arrays with practically any desirable shape [87].

Even though the effective potential $-\mathcal{R}(x,y)$ in (2) is two-dimensional, it is sufficient for the stabilization of 3D solitons. This was first suggested in combined discrete-continuous systems [88,89], and then in continuous 3D media [73]. The periodic modulation of the refractive index not only stabilizes 3D solitons (see [90] for a description of the physics of soliton formation in periodic media), but also allows controlling the relative strength of diffraction and dispersion, which is an essential ingredient in the creation of 3D states. The first observation of semi-discrete 3D light bullets, albeit in a transient state, was performed in a hexagonal fiber array with silica cores [Fig. 1(a)] illuminated by focused $170\ \mathrm{fs}$ pulses at $\lambda = 1.55\ \mu\mathrm{m}$ with peak powers up to $1\ \mathrm{MW}$ [91]. The temporal compression of such pulses results in the transient excitation of light bullets with a temporal width of $\sim 25\ \mathrm{fs}$. At such pulse duration, the Raman effect and self-steeping gradually drive the bullet below its existence threshold, making the spatiotemporal state a transient object [92]. Vortex light bullets were observed in a similar setting, also in a transient form [93]. Vortex light bullets in fiber arrays are composed of several spots (three in [93]), with the vorticity imprinted onto them (in contrast to the conventional ring spatial shape of bullets in uniform media), building a vortical structure that can coherently propagate in the array, as illustrated in Fig. 1(b).

The approach to the generation of stable 3D solitons based on the use of linear potentials, in addition to the management of diffraction, offers control over energy intervals where 3D solitons may be stable. Stable bullets can exist not only in periodic, but also in radially symmetric [Figs. 2(a) and (b)] and other types of potentials [94]. They exist in complex $\mathcal{PT}$-symmetric structures [95] too, with spatially separated and globally balanced gain and loss, where vortex bullets acquire rich shapes [Figs. 2(c) and (d)] that depend on the sign of the topological charge [96]. In such systems, the otherwise detrimental effect, introduced by linear losses, may be used to shape the beam and for controlling diffraction.

If the GVD of the material is normal, rather than anomalous, the medium supports a different type of waves, *viz.*, a nonlinear generalization of the X-wave solution of the hyperbolic Schrödinger equation [97,98]. Such waves decay in the spatial domain slower than exponentially and, to be stationary, they should carry infinite energy. Nevertheless, truncated versions were observed in waveguide arrays in 2D [99] and 3D [100] configurations. Finally, 3D linear wave packets involving combinations of Bessel and Airy shapes can also exhibit localized features, even though their invariance upon propagation in the medium is, essentially, a geometrical linear property rather than a result of self-trapping [101,102].

### 2.3. Multimode waveguides

The stabilization mechanism of multidimensional states that relies on the inhomogeneity of the refractive-index landscape has also been actively investigated in multimode optical fibers. Standard optical fibers used in telecommunications and other applications are usually designed to be single-mode, thus they may be considered as 1D media in the context of self-trapping [103]. Graded-index multimode fibers feature a large transverse area where nonlinearity can couple multiple modes, resulting in potential complex spatiotemporal field dynamics [104-110]. In particular, nonlinearity-induced locking of duration and location of relatively long pulses in different modes resulting in the formation of spatio-temporal localized states was experimentally observed in graded-index multimode optical fibers [106]. A significant contribution to the spatial confinement was provided by the linear waveguiding properties, induced by the graded-index profile of the fiber.

In this context, it is relevant to note that multiple filamentation and the generation of supercontinuum by short pulses in multimode fibers can be enhanced by the selective excitation of multiple spatial modes [107]. Spatiotemporal soliton oscillations in multimode waveguides may generate multimode dispersive waves in an ultrabroadband spectral range [108]. In the normal-GVD regime, multimode fibers were used to demonstrate nonlinearity-induced cleaning of spatial modal profiles [109,110], which is formally similar to the condensation of multimode waves occurring despite the presence of disorder, and subsequent mode decay due to spatiotemporal instabilities. Despite the complexity of multimode systems at large, they hold promising for the creation of fully 3D self-sustained states propagating over considerable distances in weakly-guiding refractive-index profiles, provided that the conditions for stable nonlinear locking of pulses propagating in different spatial modes can be achieved.

## 2.4. Exciton-polaritons

Self-trapping phenomena are a subject of intense investigation in polariton condensates that can be created in semiconductor microcavities. Polaritons are quasi-particles resulting from the coupling of cavity photons and excitons supported by quantum wells embedded into the cavity. Their low effective mass $\sim 10^{-5} m_e$ in the strong-coupling regime allows their Bose-Einstein condensation at temperatures $\sim 5\,\mathrm{K}$ [111]. Such condensates are dissipative states, therefore their persistence is supported by a resonant, or non-resonant, pump. One interesting feature of this system is the strong repulsive nonlinear interactions of polaritons through dipole-dipole repulsion of the excitonic component, which was used for the experimental observation of dark quasi-solitons and vortices [112,113] and bright 1D solitons [114,115].

The available technology of microcavity structuring has made it possible to create various lattice potentials in them [116,117], and thus to observe 1D [118] and 2D [119] gap polariton solitons (see also [121,122] for a detailed theoretical analysis). Polaritons may feature a sufficiently strong spin-orbit-coupling effect [116], produced by the energy splitting between their TE and TM polarizations. In the combination with their sensitivity to magnetic fields, the spin-orbit coupling effect opens a way for the realization of nonlinear polariton topological insulators and lasers [122,123]. Being an essentially 2D system, as the confinement in the remaining dimension is imposed by the microcavity, polaritons offer a unique platform for the experimental exploration of the physics of 2D spatial solitons in non-equilibrium condensates that can be externally controlled by the pump and by complex potential landscapes. In this connection, it is relevant to mention a very recent experimental creation of skyrmion lattice in a plasmonic field [124].

## 2.5. Plasma filaments

Multidimensional transient self-trapped states may appear as a result of filamentation of high-power laser radiation in solids or gases. Filamentation is a result of strong spatiotemporal contraction of the wave packet for powers exceeding the self-focusing threshold, leading to the formation of dynamically evolving 3D wave packets that exhibit a high degree of spatiotemporal localization [125]. The propagation of femtosecond quasi-bullets, *viz.*, transient modes which are losing energy through multiphoton ionisation, over several centimetres in fused silica was reported in [126], the formation of bullet chains due to refocusing was demonstrated in [127], and the creation of bullet filaments by single fs pulses was reported in [128]. Such 3D filaments are usually composed of a localized high-intensity core and a ring-shaped spatiotemporal periphery refuelling the central core [129]. In [130] such ring-shaped refuelling beam was used to considerably extend the propagation distance of the central filament in air, albeit in the normal-GVD regime. Such approach, when realized in a suitable material with sufficiently high and anomalous GVD, may result in the formation of long-lived 3D filaments, akin to light bullets but essentially different from ones outlined in the previous sections. New prospects for the long-range propagation of multi-terawatt femtosecond laser pulses in the single-filament regime may become available at long wavelengths (i.e., mid-infrared, in the case of air) [131].

### 3. Multidimensional states supported by nonlinear potentials

A radically different theoretical approach to the creation of stable multidimensional solitons, which has been studied in a number of settings [132-138], relies upon the use of nonlinear potentials (alias pseudopotentials [139]) induced by a defocusing nonlinearity with strength $g(\mathbf{r})$ growing fast enough from the center of the material to the periphery. The evolution of excitations in such models is governed by the modified multidimensional NLSE,

$$i\frac{\partial\psi}{\partial t}=-\frac{1}{2}\nabla^2\psi+g(\mathbf{r})|\psi|^2\,\psi. \tag{3}$$

This model predicts the existence of diverse $D$-dimensional self-trapped modes which, although decaying at $r\to\infty$, are *nonlinearizable* solutions of Eq. (3), i.e., the asymptotic form of the stationary solutions, $\psi=e^{-i\mu t}u(r)$, with real chemical potential $\mu>0$, at $r\to\infty$ is determined by the nonlinear term. In particular, for $g(r)=g_0 r^{\alpha}$, with $\alpha>0$, the asymptotic form of the solution is $u_{r\to\infty}\approx(\mu/2g_0)^{1/2}r^{-\alpha/2}$ at $r\to\infty$, for $\mu>0$. Such states can be interpreted as solitons with convergent energy, provided that $\alpha>D$, which is a fundamental condition for self-trapping in this class of models. In principle, 2D and 3D models of this type may be realizable in BECs by means of Feshbach resonances controlled by magnetic [140,141] or optical [142,143] fields with appropriate spatial profiles [144,145]. The 2D version of the model may also be realized in optics, in system where the nonlinearity of the material may be shaped by adjusting the concentration of nonlinearity-inducing dopants [146] or in hollow photonic-crystal fibers infiltrated with index-matching liquids having different nonlinearities.

Settings with spatially inhomogeneous repulsive nonlinearities of this type have been theoretically shown to support a great variety of stable 2D and, especially, 3D solitons. These may be stable vortex solitons with high values of topological charge $S$ [133,136] that tend to be highly unstable in other systems. In contrast, 3D vortex solitons predicted in this setting were found to be very robust. In particular, sudden application of a torque does not destroy them, but rather casts them in a state of stable precession, thus providing a remarkable example of a *vortex-soliton gyroscope* [136]. The model (3) also gives rise [137] to stable self-trapped toroidal vortex rings or *hopfions* [6], which carry the usual topological charge (winding number) $S$ and an intrinsic twist of the torus, characterized by an independent topological charge $m$. Illustrative examples are shown in Figs. 3(a) and 3(b). By and large, such states were considered impossible in scalar (single-field) settings. In a similar vein, hybrid 3D vortex solitons, composed of two vertically separated vortex states with equal $(S_1=S_2=1)$ or opposite $(S_1=-S_2=1)$ topological charges, may exist in anisotropic "peanut-shaped" nonlinearity landscapes [138], representing the first example of composite 3D vortex modes, illustrated in Figs. 3(c) and 3(d). Realization of all such fascinating theoretical predictions is linked to the practical feasibility of the models, which is still awaiting experimental confirmation.

### 4. Multidimensional states in atomic BECs

#### 4.1. Stabilization by optical lattices

Similar to the situation in nonlinear optics, a universal method for the stabilization of matter-wave multidimensional solitons, including vortical ones, is provided by the use of periodic potentials. These can be induced as optical lattices, i.e., interference patterns created by counter-propagating laser beams illuminating the condensate [147]. In comparison with the results in optical media, outlined above, a fundamentally novel feature for BECs is the possibility to create 3D potentials. These have been used in many experiments, such as the observation of the Mott-insulator phase [148,149], but, thus far, not yet for the creation of solitons, notwithstanding several important theoretical predictions that are available.

In particular, stable 2D gap solitons, including states with embedded vorticity, were predicted to exists in BECs with repulsive nonlinearity and a square-lattice potential more than a decade ago [150-152]. Such solitons feature unique physical properties. One of them is that 2D gap solitons are mobile, with a negative effective mass (which is a generic property of solitons of the gap type [90,147]). As a result, an additional trapping harmonic-oscillator potential superimposed on the periodic lattice expels the solitons, while the anti-trapping potential, with the inverted sign, supports stable motion along an elliptic trajectory [153]. Bright 2D solitons have also been predicted to exist in stable form in radial lattices, including radial potentials shaped like Bessel functions [154], or those represented by periodic functions of the radial variable, $\sim\cos(kr+\delta)$ [155]. Solitons created in such potentials may be strongly localized objects, capable to perform rotary motion in annular potential troughs. These and other theoretical predictions for multidimensional solitons in BECs are still waiting for experimental demonstration.

### 4.2. Stabilization by spin-orbit coupling (SOC)

Progress in elucidation of settings that may support stable 2D and 3D matter-wave solitons has been boosted by the observation of SOC in binary BECs [156,157]. A noteworthy prediction is that SOC has been theoretically shown to create the ground state in the 2D GPE with attractive nonlinearity, in the form of solitons of the SV (semi-vortex) and MM (mixed-mode) types [158-160]. In the absence of the SOC, the free-space GPE in 2D has no ground state (it is formally replaced by the collapsing solution). The evolution of binary BECs under the action of SOC is modeled by the coupled GPEs for components $\psi_\pm(x,y,t)$ of the mean-field wave function [156-160], which take the following form in 2D:

$$
\begin{aligned}
i\frac{\partial\psi_+}{\partial t} &= -\frac{1}{2}\nabla^2\psi_+ - (|\psi_+|^2+\gamma|\psi_-|^2)\psi_+ + (\lambda_{\mathrm{R}}\mathcal{D}^{(-)} - i\lambda_{\mathrm{D}}\mathcal{D}^{(+)})\psi_-,\\
i\frac{\partial\psi_-}{\partial t} &= -\frac{1}{2}\nabla^2\psi_- - (|\psi_-|^2+\gamma|\psi_+|^2)\psi_- + (\lambda_{\mathrm{R}}\mathcal{D}^{(+)} - i\lambda_{\mathrm{D}}\mathcal{D}^{(-)})\psi_+,
\end{aligned}
\tag{4}
$$

where the SOC operators are $\mathcal{D}^{(\pm)}=\partial/\partial x \pm i\partial/\partial y$, with real coefficients $\lambda_{\mathrm{R,D}}$ for the SOC terms of the Rashba and Dresselhaus types, respectively, and $\gamma$ is the relative strength of the cross-attraction between the two components. For example, solitons of the SV type with chemical potential $\mu$ are represented, in terms of the polar coordinates, by an ansatz which is compatible with Eq. (4): $\psi_+=e^{-i\mu t}\phi_+(r)$, $\psi_-=e^{-i\mu t+i\theta}r\,\phi_-(r)$, where functions $\phi_\pm(r)$ are finite at $r=0$, decaying as $\sim\exp[-(-2\mu)^{1/2}r]$ at $r\to\infty$. The SV component $\psi_-$ is the vortical one, while $\psi_+$ carries zero vorticity (hence its *semi-vortex* nature).

An essential effect introduced by SOC is the breaking of the scaling invariance of the 2D GPE, that is the source of the above-mentioned degeneracy of the TS (i.e., the single value of the norm for all TS solutions). As a result, the coupled GPEs (4) give rise to a soliton family that completely fills the interval from 0 up to norm $N_{\mathrm{T}}$ of the degenerate family of TSs, as shown in Fig. 4(a). Because soliton solutions whose norm falls below $N_{\mathrm{T}}$, which is the threshold for the onset of the critical collapse cannot initiate it, SVs becomes the ground state that is missing in the usual self-attractive GPE in 2D, as mentioned above. Furthermore, the entire SV branch satisfies the VK stability criterion, $d\mu/dN<0$. While both the SVs and MMs exist at all values of parameters, the SVs are stable at $\gamma\le 1$ and unstable at $\gamma\ge 1$, and the MMs are predicted to be stable exactly in the opposite case. The physical origin of this phenomenon arises from the fact that the SV and MM species realize the energy minimum at $\gamma\le 1$ and $\gamma\ge 1$, respectively [158]. Increasing the Dresselhaus coefficient in Eq. (4), $\lambda_{\mathrm{D}}$, while keeping $\lambda_{\mathrm{R}}=1$, eventually leads to delocalization (disappearance of solitons) at a critical point [160].

Soliton mobility in the framework of Eq. (4) is nontrivial, as SOC breaks the Galilean invariance of the usual GPE. Numerical analysis has shown that MMs exhibit mobility in one direction [$y$ in Eq. (4)], up to a critical value of the velocity, beyond which the soliton's amplitude vanishes. For the SVs, the critical velocity was found to be extremely small [158].

Regarding 3D states, SOC cannot suppress the supercritical collapse. Nevertheless, the 3D generalization of Eq. (4) has been shown numerically to give rise to 3D *metastable* soliton states, which realize local energy minima, being therefore are stable against small perturbations [161]. As well as in the 2D geometries, the 3D system gives rise to metastable solitons of the SV and MM types at $\gamma \leq 1$ and $\gamma \geq 1$, respectively, as illustrated in Figs. 4(b) and 4(c). All results described in this section are awaiting experimental exploration, which may be relatively easy to implement.

### 4.3. Giant vortex solitons in binary BECs

Theoretical efforts have been devoted to elucidating a physically realizable BEC model that may support stable vortex solitons with high values of vorticity $S$. Note that, in SOC systems, all such states have been shown to be unstable [158]. A relevant 2D model, which produces stable vortex solitons with high values of $S$ was put forward in [162], in the form of coupled GPEs for wave functions representing two different hyperfine atomic states in an ultracold bosonic gas, resonantly coupled by a magnetic component, $H$, of a microwave electromagnetic field. In turn, the microwave field is generated by the Poisson equation with the respective source density, $\psi_+\psi_-^*$. Eliminating the magnetic field by means of the Green function of the 2D Poisson equation, the corresponding governing equation can be cast in the form featuring an effectively nonlocal interaction (possibly combined with the local nonlinearity):

$$i\frac{\partial \psi_\pm}{\partial t} = -\frac{1}{2}\nabla^2\psi_\pm - \beta|\psi_\pm|^2\psi_\pm + \Gamma\psi_\mp \int \psi_\pm(\mathbf{r}')\psi_\mp^*(\mathbf{r}')\ln|\mathbf{r}-\mathbf{r}'|\,d\mathbf{r}', \quad (5)$$

where $\Gamma$ is the strength of the feedback of the wave function onto the magnetic field, and $\beta$ is the strength of the contact self-interaction of the BEC. Solving Eq. (5) numerically produces vortex solitons which are predicted to be stable, at least, up to $S=5$ (hence the term "giant vortices") [162]. Illustrative examples are shown in Fig. 5. Similar to the above-mentioned exciton-polariton system, these are hybrid solitons, as they include both the matter-wave and field components. Their experimental observation is still an open question.

### 4.4. Stable quantum droplets in binary BECs

We now turn to a recent breakthrough achievement that opens up a whole new domain in the area of multidimensional self-trapping. This is the creation of a new class of quantum liquids [20-25], which are fully coherent and extraordinarily diluted (their density is of the order of $\sim 10^{14}$ atoms/cm$^3$, i.e., eight orders of magnitude smaller than that of liquid helium [20]). Bulgac [163] proposed already in 2002 that bosonic droplets (or "boselets") could be created if the mean-field (MF) energy of self-attraction is compensated by three-body repulsive terms, resulting in a minimum of the energy as a function of density. Alternatively, a few years ago Petrov showed theoretically that the repulsive Lee-Hung-Yang (LHY) corrections [16], induced by quantum fluctuations around the MF state, can be exploited for the stabilization of the two-component condensates [17,18] (see also [164]). In 3D geometries, the interactions should be tuned in such a way that the self-repulsion in each component is slightly overbalanced by attraction between the components [17]. Note that the strength of interactions can be tuned by exploiting a Feshbach resonance. While such systems would be unstable in models based only in the MF approximation, the LHY corrections were predicted to make them stable.

Unlike classical liquids, in which the average density takes values on the order of the interaction potential (typically of the Van der Waals form), in quantum droplets it is controlled by the balance of the MF and LHY terms, and thus can be tuned. The diluteness of the quantum droplets justifies the applicability of the perturbation theory, in contrast to usual quantum liquids, where such approach is not valid. The separation of scales of the underlying soft and hard modes allows the derivation of the LHY terms in the local form, depending solely on the density. For a mixed binary fluid with equal wave functions $\psi$ of the two components, this implies addition of a self-repulsive quartic term with coefficient $g_{\mathrm{LHY}}>0$ to the 3D GPE with the effective MF cubic self-attraction, with respective coefficient $g>0$ [17], as:

$$i\frac{\partial\psi}{\partial t}=-\frac{1}{2}\nabla^{2}\psi-g\left|\psi\right|^{2}\psi+g_{\mathrm{LHY}}\left|\psi\right|^{3}\psi. \tag{6}$$

Exact Monte-Carlo calculations [18,167,168] and approximate variational estimates [169] have been found to agree with the predictions of Eq. (6) in the limit of weak interactions. Modifications of the system to address Rabi-coupled [165] and SOC-coupled [26] mixtures, as well as Bose-Fermi systems [166], were recently elaborated too. Reduction of the 3D model (6) to lower dimensions, under the action of tight confinement, has been presented in [18]. In the same case of equal wave functions of the two components as implied in Eq. (6), the 2D model amounts to the single GPE with a nonlinear term $\sim\left|\psi\right|^{2}\psi\ln\left|\psi\right|^{2}$, which implies self-attraction for small densities $\left|\psi\right|^{2}$, and repulsion for large $\left|\psi\right|^{2}$. Such nonlinearity has been theoretically shown to give rise to stable standard solitons [18,26]. The 1D geometry is special in that the beyond-MF correction turns out to be effectively attractive, $\sim-\left|\psi\right|\psi$, thus the quantum liquid is formed when the MF term is slightly repulsive [18,170].

Ultradilute quantum droplets have been realized experimentally recently [20-22]. In single-component dipolar condensates [19,171-177], the short-range interactions were tuned in such a way that the MF energy, including contributions stemming from the short-range and anisotropic long-range dipolar potentials, were slightly attractive, while the repulsive LHY correction stabilized the system [174,175] against collapse [176]. On the other hand, unlike fundamental dipolar quantum droplets, their counterparts with embedded vorticity were found to be unstable [177].

Stable single-component dipolar quantum droplets were created recently in condensates of $^{164}$Dy [20,21] and $^{166}$Er [22]. It was experimentally verified that the stabilization mechanism indeed originates from quantum fluctuations [21]. Due to the nature of the dipole-dipole interactions, the dipolar quantum droplets feature strong anisotropy, as shown by the 3D simulations displayed in Fig. 6(a). Creation of 3D states in gases with contact interactions was recently demonstrated experimentally [23-25], using binary mixtures of two hyperfine atomic states of $^{39}$K. In these experiments, the necessary balance between the repulsive and attractive intra- and inter-component interactions was achieved by means of the Feshbach resonance. The experiments were performed both in the presence of a strong confining potential, applied in one direction, which leads to the formation of quasi-2D states [23,24], and in the absence of the confinement, which allowed the observation of isotropic droplets [25]. Both, quantum droplets and standard self-trapped states, which may be understood as quantum-gas solitons, were created, and the boundary between the quantum droplet regime and the standard soliton regime was mapped [24].

A noteworthy recent theoretical result is the prediction of stable quantum droplets with embedded vorticity [178]. The corresponding solution to Eq. (6) with chemical potential $\mu$ and integer vorticity $S$ has the form of $\psi=\exp(-i\mu t+iS\theta)\phi(\rho,z)$ in cylindrical coordinates $(z,\rho,\theta)$, where the real function $\phi(\rho,z)$ decays exponentially at $\rho\rightarrow\infty$ and $\left|z\right|\rightarrow\infty$, and vanishes as $\sim\rho^{|S|}$ at $\rho\rightarrow0$. The family of vortical states with $S=1$ was found to be partly stable, as shown in Fig. 7(a). A narrow stability region was also found to exist for $S=2$, as depicted in Fig. 7(b). Estimates indicate that the experimental creation of such states in a mixed condensate $^{87}$Rb-$^{41}$K requires densities of the order of $\sim 5\times 10^{15}$ atoms/cm³. Reaching such states experimentally is challenging at present, but they should become achievable as the state of the art advances.

## 5. Multidimensional solitons in liquid crystals and liquid ferromagnets

Another field in which the experimental efforts have recently produced remarkable results is the study of self-sustained states in static material structures (rather than propagating fields in optical media, or matter-waves in BECs) in liquid crystals and in ferrofluids, made of colloidal suspensions of disk-shaped magnetic nanoparticles [179-181]. The aim of the experiments is the creation of topologically structured 3D solitons, such as the above-mentioned hopfions [an example is shown in Fig. 8(a)] and doughnut-shaped *torons*, i.e., a twisted cylinder closed on itself in the form of a torus, similar to the above-mentioned hopfions, and coupled to a surrounding uniform field [illustrated in Figs. 8(b) and 8(c)]. The nature of these systems and 3D states which they support are quite different from those addressed in previous sections.

In nematic liquid crystals, hopfions are realized as finite-energy configurations of the corresponding order parameter (director field), characterized by an intrinsic knotted structure, with different particular knotted configurations identified by values of the corresponding Hopf topological number, which represents different types of states [179]. Each particular type may be characterized by the so-called *pre-image*, i.e., a closed loop in the 3D space, which exhibits the same knottedness as carried by the respective hopfion in the liquid-crystal medium. In [179], various species of hopfions were created experimentally, using laser tweezers to set the necessary configurations of the order parameter, and an optical imaging technique was employed to observe the 3D shapes of the hopfions. The actual size of the modes created in the nematic samples is measured on the micron scale. Thus, pre-images of these 3D hopfion modes were reconstructed from the experimental data, and, in parallel, they were reproduced numerically, using the Hamiltonian of the nematic liquid crystals, expressed in terms of the local order parameter. It is essential to stress that, as stated by the Hobart-Derrick theorem [182,183], rigorous energy arguments demonstrate that 3D solitons, built of a single real scalar field cannot be stable, unless the underlying Hamiltonian density includes higher-order spatial derivatives, in addition to the standard squared-gradient term. The hopfions found in [179] may be stable because they are based on a vectorial order parameter.

Similar experimental and theoretical results were reported in [181] for hopfions experimentally created in liquid-colloidal ferromagnets, in which the same role as the local director in the above-mentioned liquid crystals is played by the local direction of the magnetization (while its absolute value is fixed), the respective Hamiltonian being similar to its liquid-crystal counterpart. In this case too, pre-images of the experimentally created hopfions were produced experimentally and reproduced, in parallel, by numerical computations using the underlying Hamiltonian expressed in terms of the local magnetization.

It is relevant to mention the recent experimental creation of a 3D stable *nontopological* soliton, in the form of a confined region of oscillating nematic director. It propagates in the liquid crystal at a high velocity, in the presence of dc electric filed, perpendicular to the field and to the original orientation of the director [184].

Despite the rather specific nature of these systems, in the spirit of the present review, we stress their importance as a way to experimentally construct distinct species of fully macroscopic and perfectly stable self-trapped topologically organized states under well-controlled conditions. Further, the use of well-established underlying Hamiltonians makes it possible to accurately identify the observed modes as expected complex topological states. To illustrate cross-fertilization between different areas that their mutual similarities may boost, it is instructive to mention that a possibility of the existence of 3D structures defined by knotted core lines of the mean-field wave-function patterns, was recently predicted for BEC loaded into an optical lattice [185]. Their observation and generalization to other topologies are open questions.

## 6. Conclusion and outlook

We have presented a concise overview of the progress in the creation of dynamically stable multidimensional soliton-like states, with a focus on optical materials, matter-wave condensates, ultradilute quantum liquids related to them, as well as liquid crystals and ferrofluids. However, it has to be properly appreciated that the topic is much broader as it encompasses many areas of physics, therefore the central goal of this work is reaching different communities, in the expectation of cross-fertilization due to interplay between them. The motivation is taming the unique opportunity offered by multidimensional geometries for the creation of highly localized self-sustained states with rich internal structure, including topological states such as vortex tori, hopfions, skyrmions, vortex-antivortex complexes, vortex-soliton gyroscopes and more, which, by their very nature, are not possible in 1D geometries.

The main conclusion to be drawn from the intense efforts conducted in the above-mentioned areas during more than three decades is that theoretical predictions are far more advanced and numerous than experimental demonstrations, especially in the case of long-lived, fully 3D self-sustained states. Various 2D states (both spatial and spatio-temporal ones) have been observed experimentally in different systems, but fully 3D states have proven to be much harder to create. The main challenge, which remains to be overcome in most settings, is the ubiquitous trend of multidimensional states, both in 2D and even more so in 3D, to be strongly unstable in most available natural materials. Salient exceptions are known to theoretically exist, such as settings based on materials with quadratic nonlinearities, which are readily available even in commercial devices, but where the required conditions to generate 3D localized states have never been met to date.

The above is in sharp contrast to 1D solitary waves, which exist as stable and robust states in an astonishing, and continuously growing, variety of physical systems supported by different mechanisms, and that are not only fascinating scientific objects but, in some cases, they are also essential to practical applications. A nascent example is the use of dissipative Kerr temporal solitons for the generation of robust frequency combs in compact microresonators that are of paramount importance to high-precision techniques in a wide variety of scientific and technological areas [186].

Various theoretical methods for the stabilization of multidimensional self-sustained states against the collapse, self-breaking and azimuthal self-splitting (the latter problem being crucially important to vortex states) have been put forward, the most straightforward ones being the use of non-cubic nonlinearities, such as saturable, quadratic, competing or synthetic nonlinearities. Another stabilization method has been theoretically shown to be offered by spatially periodic potentials in the form of optical lattices. Such approaches hold both for optical solitons and for standard BECs where even fully 2D stable states have not yet been observed. In this regard, new promising stabilization methods have been recently theoretically predicted in two-component BECs. In particular, the spin-orbit coupling between the two components has been theoretically shown to afford the creation of completely stable 2D solitons and metastable 3D ones.

A very recent remarkable breakthrough in the field is the experimental creation of quantum droplets, which are made completely stable in 2D and also in 3D geometries by the effects of quantum fluctuations in binary condensates. The fluctuations are represented by the Lee-Hung-Yang corrections to the mean-field Gross-Pitaevskii evolution equation. In these ultradilute quantum liquids, self-trapping is provided by the attractive interaction between the two components (or by dipole-dipole interactions in the single-component condensate), competing with the local self-repulsion in each component. Beyond their intrinsic interest, these observations open up an important direction for exploration of fundamental and potentially widely-applicable concepts directly linked to multidimensionality. In particular, a fascinating future direction of the research in this context is the possibility to create topologically structured varieties of 2D and 3D quantum droplets, the existence of the simplest of which has been already predicted theoretically. The conditions required for their experimental creation and observation, in particular as concerns the droplets' sizes and density, are challenging to meet in currently available condensates, but no fundamental obstacles that may prevent their realization in the future are foreseen, as the state of the art advances.

A final comment is in order: During the last years, optical settings and BECs held in optical lattices have been explicitly shown to provide powerful tools to mimic the behavior of other physical systems and phenomena – chiefly, these are fundamental effects in condensed matter, topological, and quantum systems – that may be much harder to access experimentally in the original settings, such as spin-orbit coupling. Formation of multidimensional self-trapped states in the corresponding systems should be seen from a similar perspective, hence their fundamental, broad, and far-reaching importance.

**Acknowledgments**

We greatly appreciate many valuable collaborations and discussions with Sadhan K. Adhikari, Guangjiong Dong, Arnaldo Gammal, Randall G. Hulet, Vladimir V. Konotop, Oleg D. Lavrentovich, Yongyao Li, Dumitru Mihalache, Dmitry S. Petrov, Hidetsugu Sakaguchi, Luca Salasnich, Evgeny Ya. Sherman, Yasha Shnir, Dmitry V. Skryabin, and Leticia Tarruell.

L.T. and Y.V.K. acknowledge support from the Severo Ochoa program (SEV-2015-0522) of the Government of Spain, Fundacio Cellex, Fundació Mir-Puig, Generalitat de Catalunya and CERCA. The work of B.A.M. is supported, in part, by the joint program in physics between NSF and Binational (US-Israel) Science Foundation through project No. 2015616, and by Israel Science Foundation through Grant No. 1286/17. This author appreciates hospitality of ICFO during the preparation of this review. G.E.A. acknowledges financial support from the MICINN (Spain) Grant No FIS2017-84114-C2-1-P.

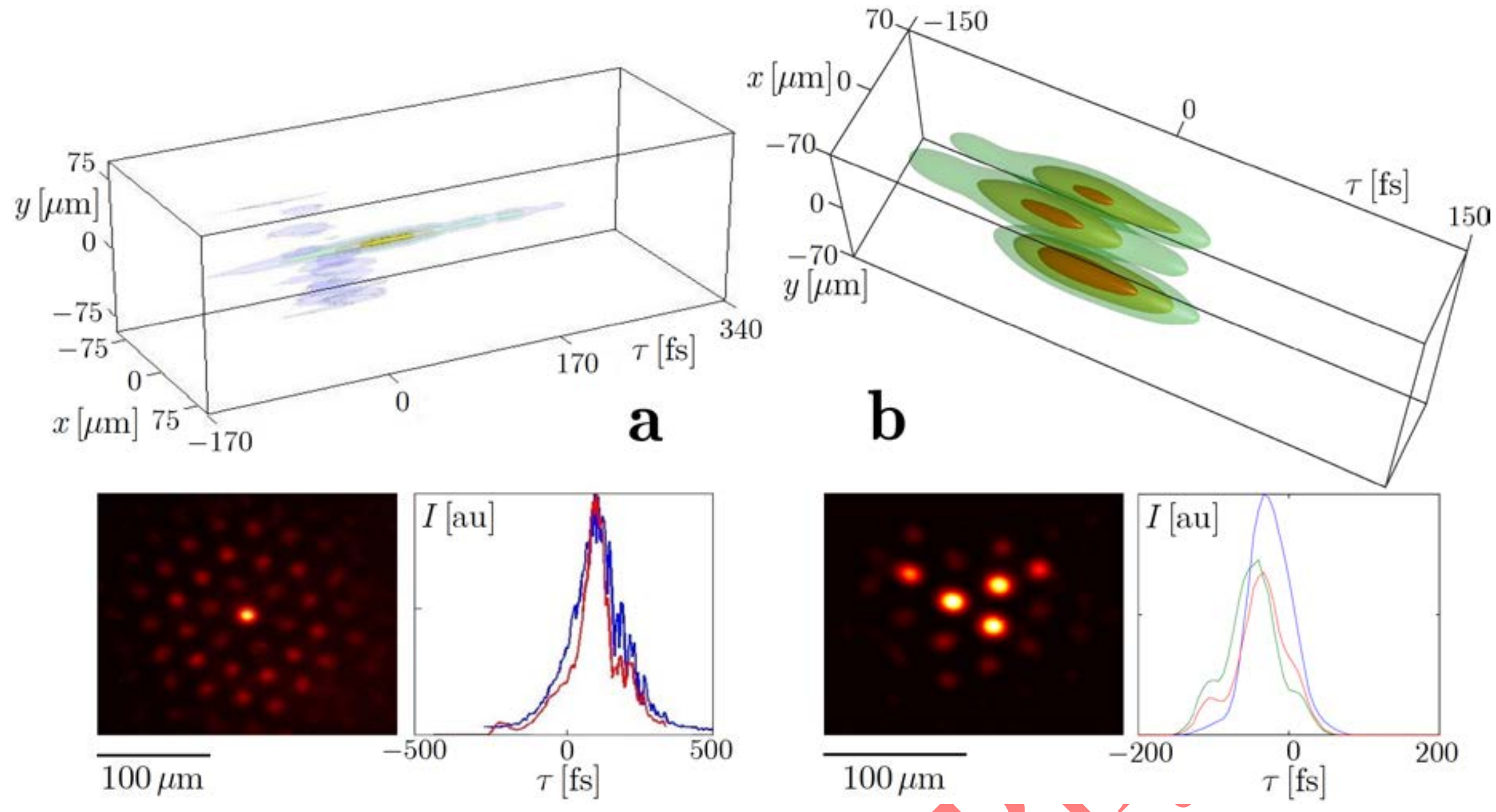

**Fig. 1. Observed excitation of short-lived fundamental and vortex light bullets in hexagonal waveguiding arrays**. The top row shows isointensity levels of the experimentally measured spatiotemporal structure of an excited fundamental light bullet centered on a single waveguide [91] **(a)**, and a vortex light bullet occupying three waveguides [93], as reported by Minardi and co-workers **(b)**. The bottom row shows the corresponding time-integrated output intensity distributions illustrating the spatial structure of the bullet (left images), and temporal cross-correlation traces in the central waveguide for the fundamental light bullet, or in three excited waveguides for the vortex bullet (right images). The overlap of traces in the latter case confirms that wave packets forming the bullet propagate synchronously. Bullets were excited at peak powers $\sim 0.5$ MW **(a)** and $\sim 1.6$ MW **(b)** with femtosecond pulses. Thus, in both cases, higher-order effects led to their decay.

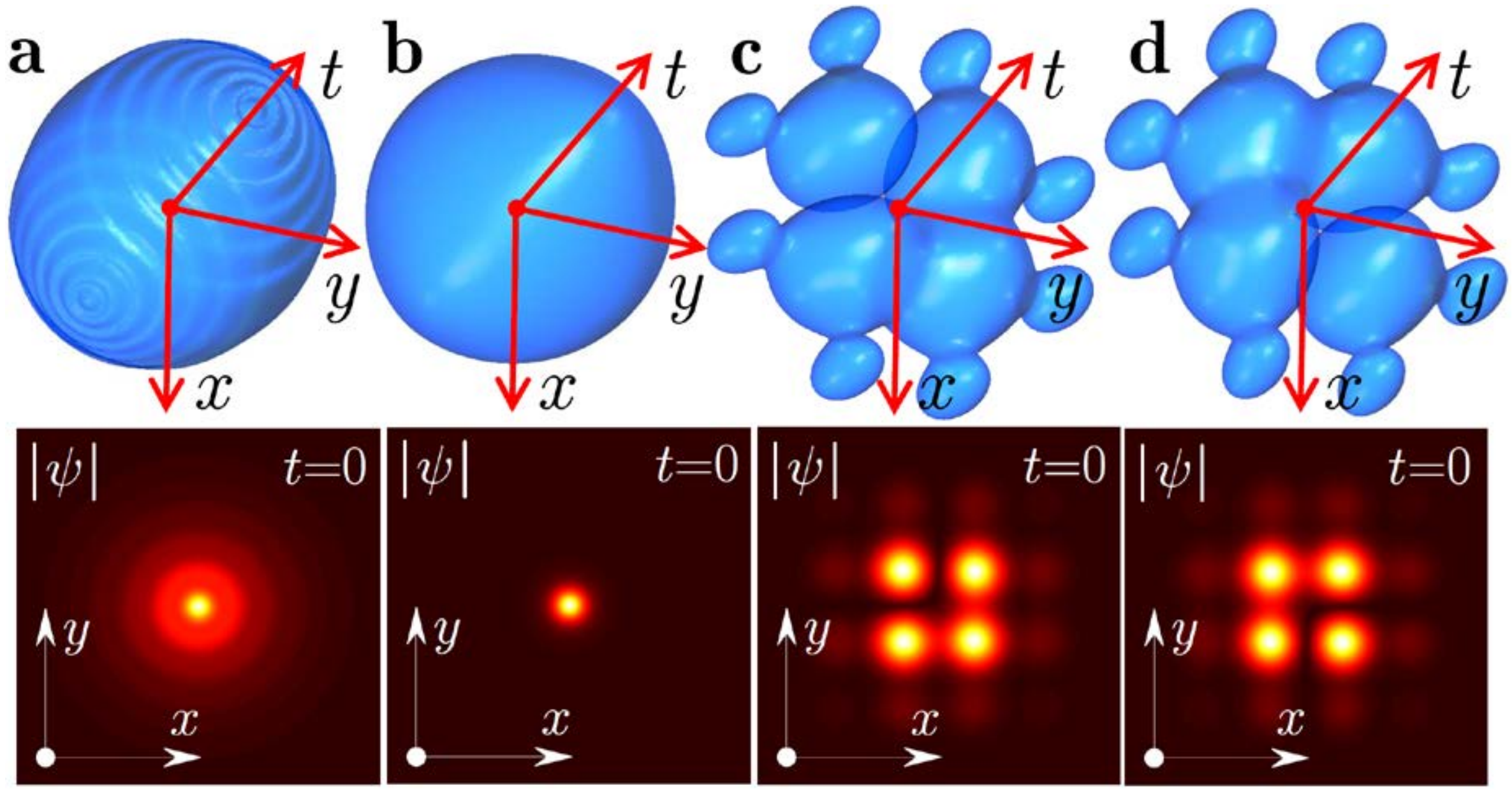

**Fig. 2. Predicted light bullets in radially-symmetric and complex potentials**. Isosurface plots (top) and spatial distributions of the field's absolute value at $t=0$ (bottom) show extended **(a)** and strongly localized **(b)** stable fundamental bullets supported by the two-dimensional Bessel lattice, as per [94], as well as vortex bullets with topological charges $+1$ **(c)** and $-1$ **(d)**, supported by a two-dimensional periodic $\mathcal{PT}$-symmetric lattice, as per [96]. Note the symmetry imposed by the lattice potential on the spatial profile of the bullets. In the $\mathcal{PT}$-symmetric lattice, the shape of the bullet depends on the sign of its topological charge.

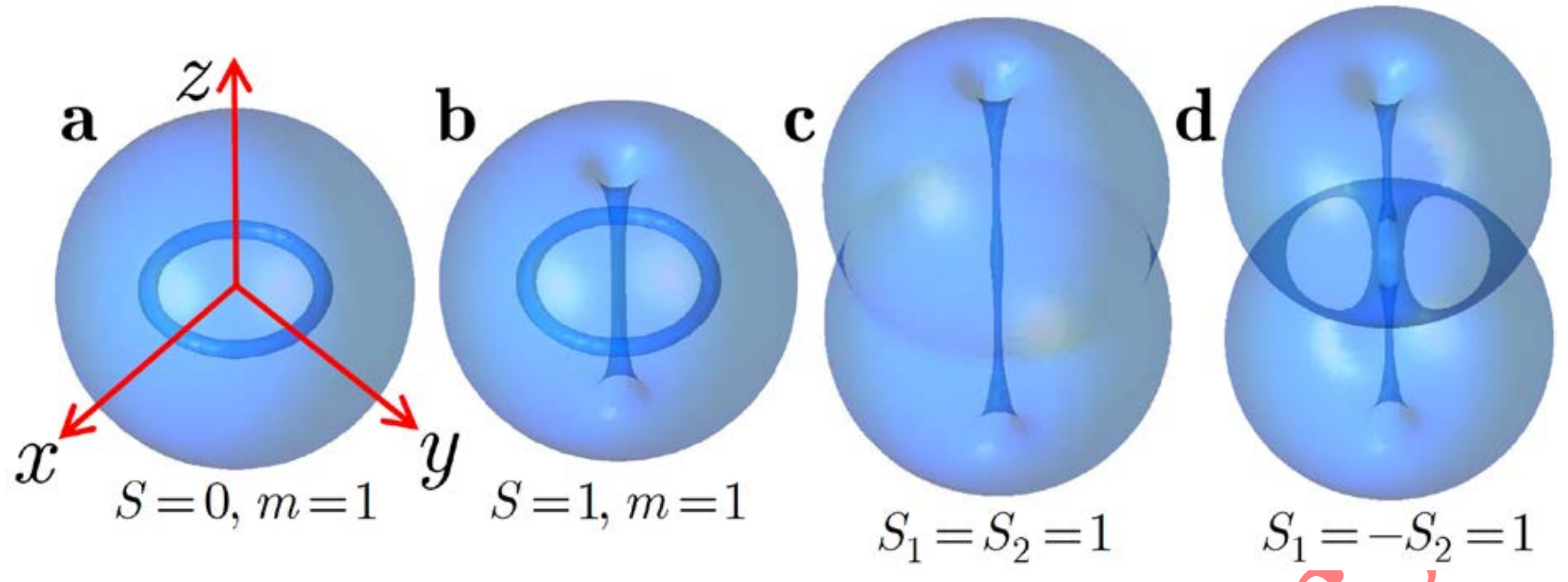

**Fig. 3. Predicted Hopfions and hybrid 3D vortex solitons.** Theoretically-envisioned inhomogeneous repulsive-nonlinearity landscapes provide rich possibilities for confinement and stabilization of 3D solitons that are either unstable or do not even exist in other settings. Examples shown here by isosurface density plots include stable hopfions, i.e., twisted tori with vorticity $S=0$ **(a)** and $1$ **(b)**, and the intrinsic twist of the torus with topological number $m=1$ in the model with the anti-Gaussian nonlinearity profile, as per [137], as well as stable hybrid vortex solitons, whose upper and lower parts can carry equal $S_1=S_2=1$ **(c)** or opposite $S_1=-S_2=1$ vorticities **(d)**, in a model where the repulsive nonlinearity grows at $r\to\infty$, while having two local minima on the $z$ axis, as per [138].

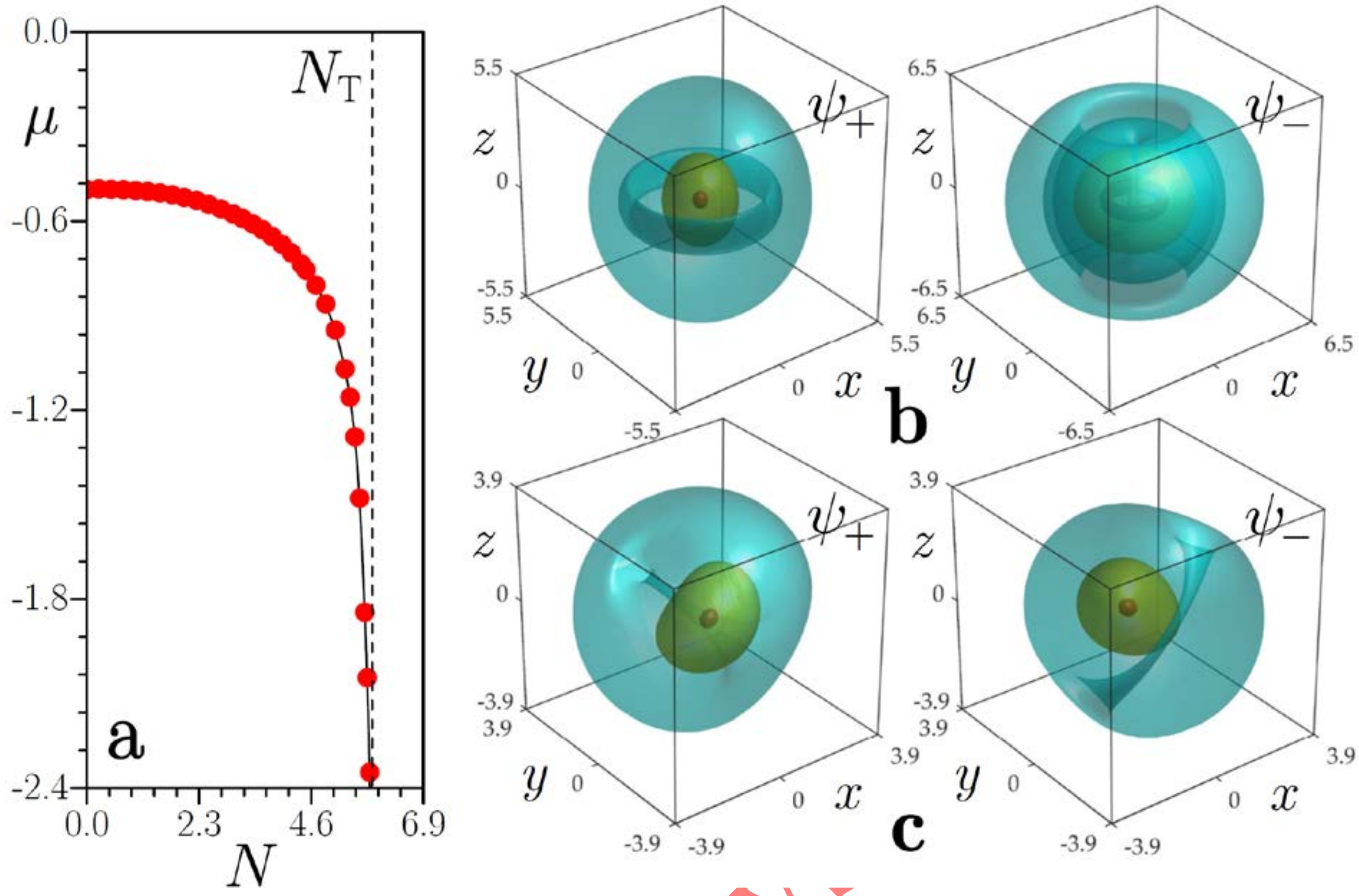

**Fig. 4. Predicted stable multidimensional solitons in BECs with spin-orbit coupling**. Panel (a) shows the chemical potential versus norm for the 2D SV solitons predicted to exist in the model (5) with $\lambda_R = 1$, $\lambda_D = 0$, $\gamma = 0$, as per [158]. The entire family is predicted to be stable in accordance with the VK stability criterion. The limit value $N = N_T$ at $\mu \to -\infty$ is the norm of the degenerate TS family. In panels (b) and (c), sets of three isosurface density plots display examples of 3D metastable solitons of SV (b) and MM (c) types in BECs with SOC, as per [161].

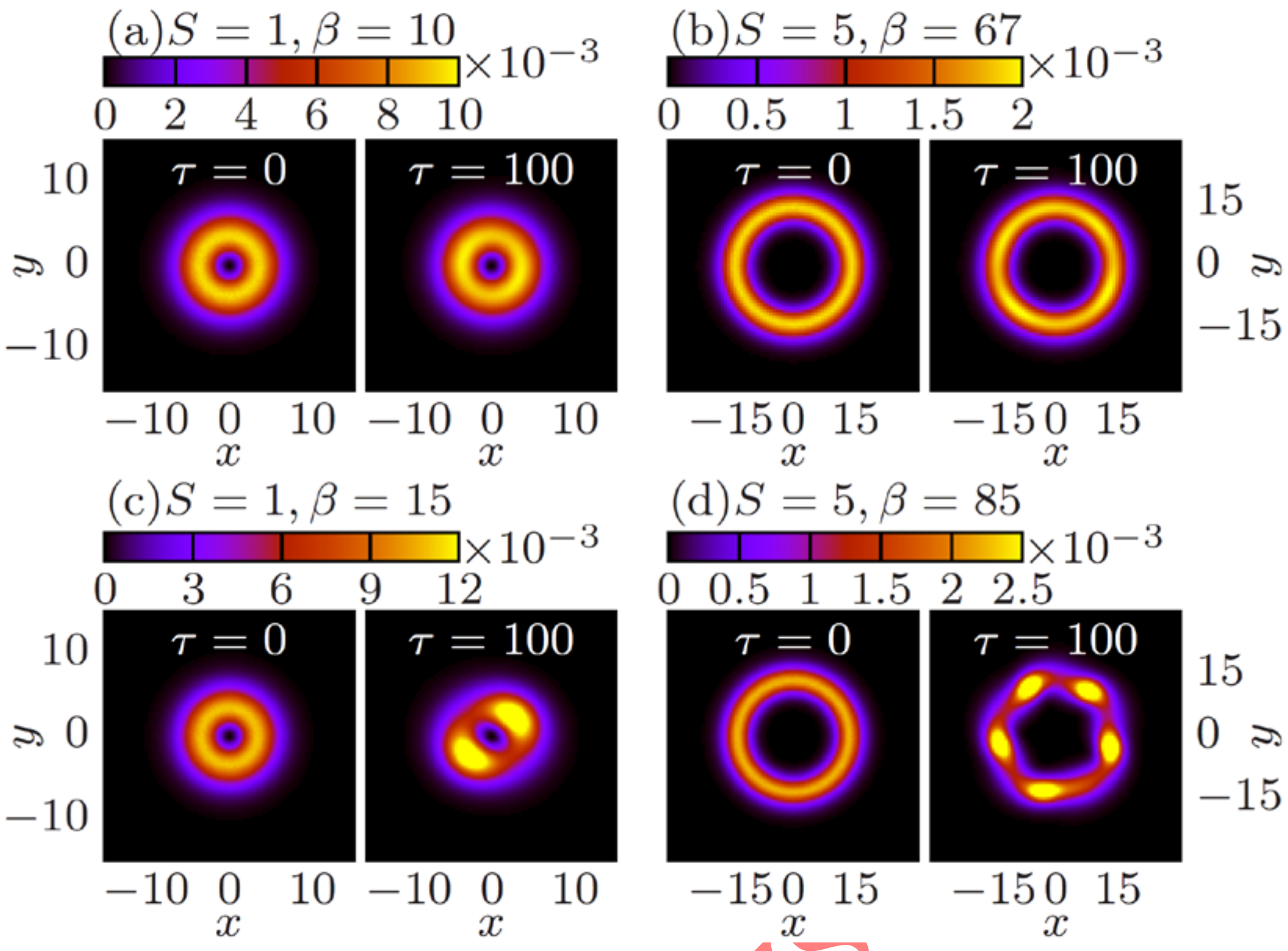

**Fig. 5. Predicted vortex solitons in the binary BECs coupled by a microwave field.** Stable (top) and unstable (bottom) evolution of the hybrid vortex solitons with topological charges $S=1$ and $5$. The unstable solutions are predicted to split into necklace-like patterns, which may keep their shape for a while in the course of the subsequent evolution [162].

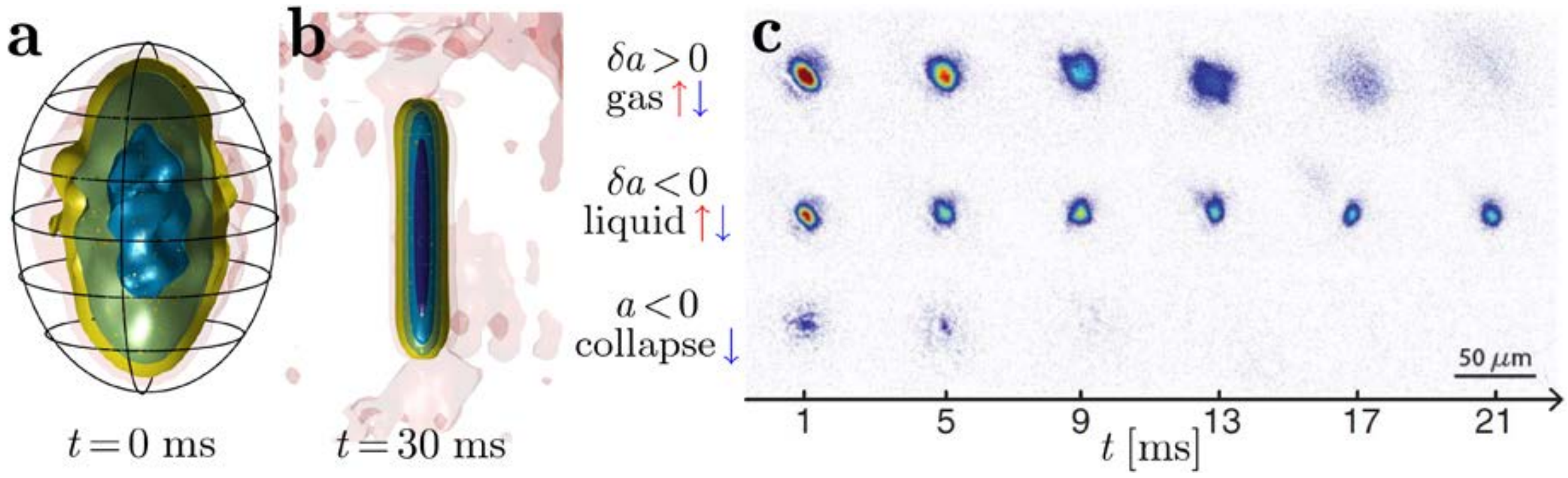

**Fig. 6. Predicted and observed self-sustained multidimensional quantum droplets.** Dipolar droplets that were predicted to form in $^{164}$Dy atoms [19] before **(a)** and after **(b)** turning off the trapping potential and adjusting the $s$-wave scattering length. The black lines show the shape of the external trap when it is present. **(c)** Experimental observation of quantum droplets in a two-component condensate of $^{39}$K atoms [23]: expansion of a two-component gas (top row), stable self-bound evolution of two-component liquid (middle row) and collapse of a single attractive component (bottom row). The interplay between the MF interactions and quantum fluctuations stabilizes the droplets.

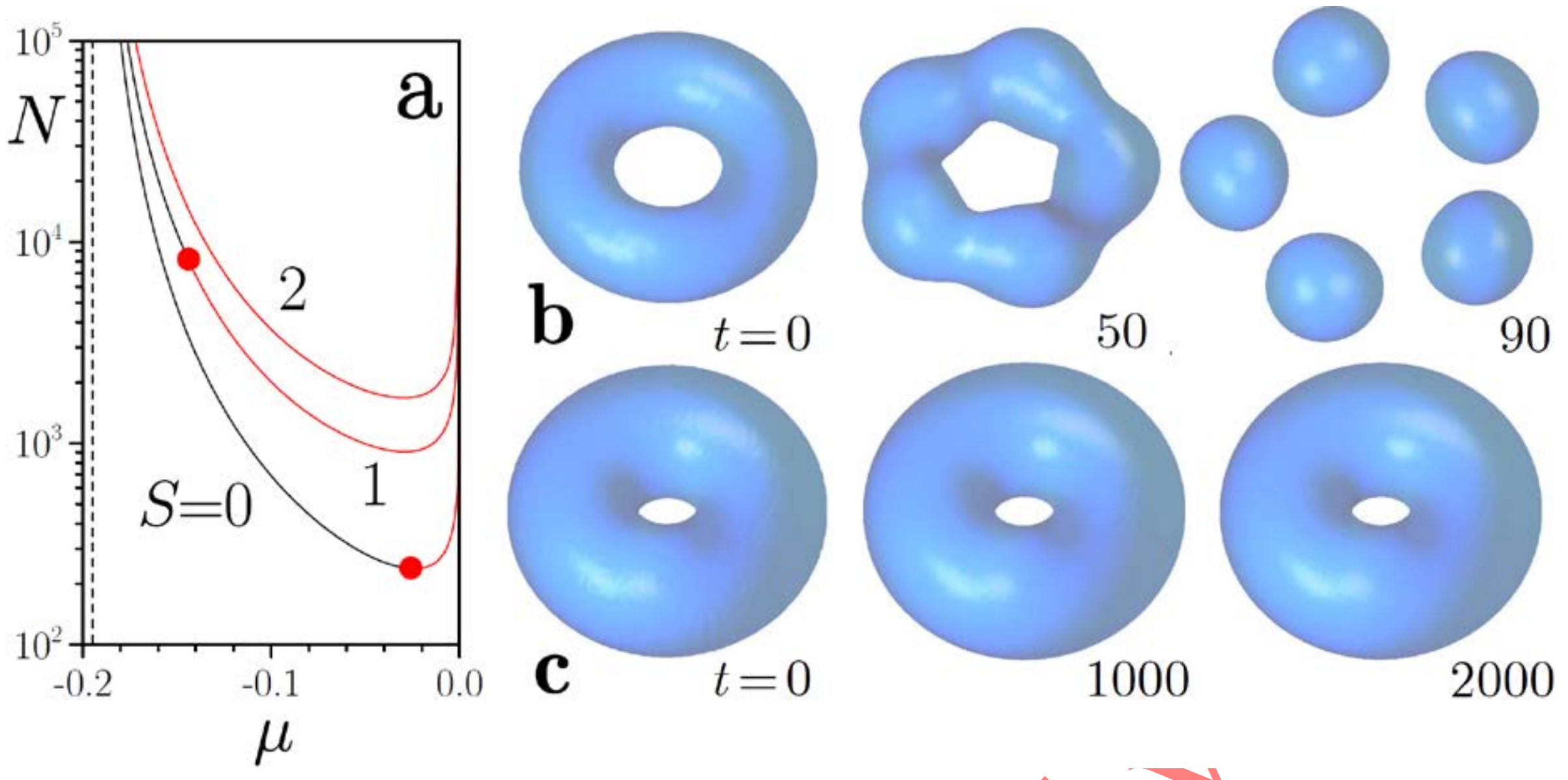

**Fig. 7. Predicted stable quantum droplets with embedded vorticity. (a)** The norm of states with vorticities $m_{1,2} \equiv S$ in the two components versus chemical potential. Red and black segments, separated by red dots, correspond to unstable and stable states, respectively. For $S = 2$, a stability segment exists too, but it is located at large values of the norm outside of the region displayed here. Examples of the evolution of unstable **(b)** and stable **(c)** vortical states with $S = 2$ and $\mu = -0.04$ and $\mu = -0.183$, respectively. The results are displayed for $g = 1.75$ and $g_{\text{LHY}} = 0.50$, as per [178].

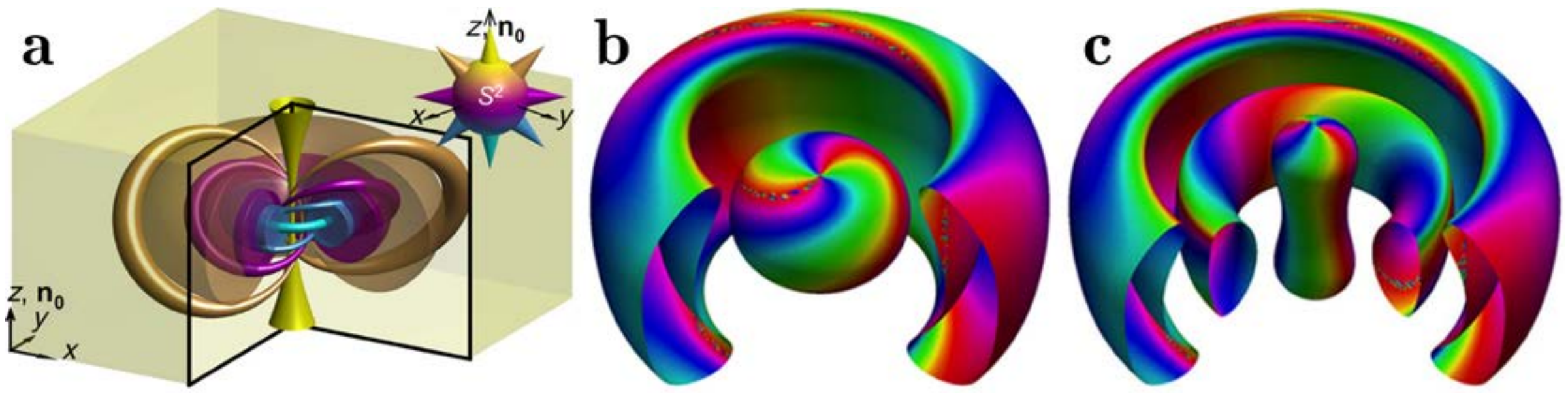

**Fig. 8. Observed topological solitons with differently knotted nematic fields in a liquid crystal. (a)** A closed-loop preimage (regions in the 3D liquid-crystal sample that have the same orientation of the physical field), and isosurfaces drawn at $n_z = 0$, illustrating the structure of 3D torons with complex linking, and $3\pi$ **(b)** or $5\pi$ **(c)** twist between their central axes and the far-field periphery, as per [179]. Surface colors denote different azimuthal orientations in the $(x, y)$ plane of the nematic director $\mathbf{n}(\mathbf{r})$ in the topological soliton.